\documentclass[%
 reprint,
 amsmath,amssymb,
 aps,
floatfix,
]{revtex4-1}

\usepackage{graphicx}

\graphicspath{{suman/}}
\usepackage{dcolumn}
\usepackage{bm}
\usepackage{hyperref, graphicx}

\begin{document}

\preprint{APS/123-QED}

\title{Dynamical Properties and Effects of Quantum Diffraction on The Propagation of E-A-Solitary Waves in Three-Component Fermi Plasma}


\author{Tamal Ghosh}
\email{tamalghosh695@gmail.com}
 \affiliation{Department of Physics, Bidhannagar Govt. College, Kolkata, West Bengal 700064, India}
\author{Suman Pramanick}%
 \email{sumanphysics@kgpian.iitkgp.ac.in}
\affiliation{Department of Physics, Indian Institute of Technology Kharagpur, Kharagpur, West Bengal 721302, India}

\author{Soumya Sarkar}
\email{sarkarsoumya65@gmail.com}
\affiliation{Department of Physics, National Institute of Technology Karnataka, Karnataka 575025, India}%

\author{Ankita Dey}
\email{ankitaankidey@hotmail.com}
\affiliation{Department of Physics, Lady Brabourne College, Kolkata, West Bengal 700017, India}%

\author{Swarniv Chandra}
\affiliation{Govt. General Degree College at Kushmandi, Dakshin Dinajpur, 733121, India}



             
\begin{abstract}
Electron Acoustic Solitary structures in Fermi Plasma with two temperature electrons have various applications in space and laboratory-made plasma. Formulation of an adequate theory is important to understand various physical systems with various physical parameters. The motion of two temperature electrons in a quantum Fermi plasma system highly affects the solitary profile of the system. We study the quantum Fermi plasma system with two temperature electrons where the streaming velocities of the two-electron population are opposite. We consider the quantum hydrodynamic model (QHD) and derive a linear dispersion relation for the system. For the non-linear study of the system, we use the standard perturbation technique to derive KdV-B equation and show the evolution of a solitary profile with different plasma parameters. We analyze the stable Rouge wave structure using NLSE and show simulation results. We study the dynamical properties and phase plot for a two-stream quantum Fermi plasma system with two temperature electrons.
\end{abstract}


\maketitle


\section{Introduction}\label{intro}
Plasma systems containing two distinct groups of electrons shows Electron acoustic waves (EAWs). The distinction between two electron groups comes from their energy. Two types of electrons are 1) Hot electrons and 2) Cold electrons. The frequency of electron acoustic mode is higher than ion acoustic frequencies of plasmas. Hot electrons can freely move with less viscous drag and supply restoring force, whereas cold electron feels viscous drag and produce inertia to the system. The thermal speed of hot electron is very large in comparison to cold electron. The phase speed of electron acoustic wave is smaller than the thermal speed of hot electrons but much larger than thermal speed of cold electrons. In this system ions may be considered as uniform neutralizing background. EAWs with two groups of electrons plays an important role in space plasma (Ang and Zhang (2007)\cite{ref17}; Barnes et al. (2003)\cite{ref20}; Feldman et al. (1975)\cite{ref30}, (1983a)\cite{ref31} as well as laboratory made plasmas (Ditmire et al. (1998)\cite{ref27}; Sheridan et al. (1991)\cite{ref50}; Armstrong et al. (1979)\cite{ref18}; Kadomtsev and Pogutse (1971)\cite{ref37}; Henry and Trguier (1972)\cite{ref36}; Defler and Simonen (1969)\cite{ref25}). The source of broadband electrostatic noises can be addressed with EAWs and it has been used to explain those. The wave emission in different regions of earth’s atmosphere can be explained with EAWs. For these studies EAWs has become one of the important research areas in plasma physics. In recent years there is a boost in the study of the nonlinear evolution of EAWs (Bains et al. (2011)\cite{ref19}; Soultana and Kourakis (2011)\cite{ref55}; Kourakis and Shukla (2004)\cite{ref39}; Singh and Lakhina (2001)\cite{ref54}). Various space-craft missions, e,g, the FAST at the auroral region (Ergun et al. (1998a\cite{ref28}, 1998b\cite{ref29}); Delory et al. (1998)\cite{ref26}; Pottelette et al. (1999)\cite{ref46}) and the POLAR and GEOTAIL missions in the magnetosphere (Matsumoto et al. (1994)\cite{ref44}; Franz et al. (1998)\cite{ref33}; Cattell et al. (2003)\cite{ref23}) explained by EAW related structures. Most of these application sites need theory and better understanding of non-relativistic classical plasmas. However, there are numbers of works on the theory of nonlinear propagation of electrostatic modes in quantum plasmas with consideration of quantum hydrodynamic model of plasma (Shukla and Eliasson (2006)\cite{ref52}; Sahu and Roychoudhury (2006)\cite{ref48}; Ali and Shukla (2006)\cite{ref51}; Manfredi (2005)\cite{ref41}; Haas et al. (2003)\cite{ref35}; Gardner and Ringhofer (1996)\cite{ref40}). The non-linear wave structure of cold and hot electrons are investigated \cite{ref1}-\cite{ref7}

In this paper we study linear and nonlinear properties of EAWs in Fermi plasma with two temperature electrons. We first assume basic hydrodynamic equations for the system then we normalize them using suitable scaling for the system. With the normalized equations on hand we linearize them to get linear dispersion relation and linear dispersion characteristics are investigated. The Korteweg-de Vries Burgers (KdV-B) equation is derived using standard perturbation technique. We investigate dependence of soliton properties on different plasma parameters. We study Rouge wave formation for the system. We investigate the dynamical nature of the system. Then we conclude the paper with some remarks and future work plans.

\section{Basic Formulation}\label{basic}
The plasma system that we considered is
unmagnetized consisting of two groups of electrons.
Two groups are 1) Hot electrons and 2) Cold
electrons. The thermal energy of hot electron is
higher than cold electrons so the mobility of hot
electron is large compared to that of cold electrons.
So, in the momentum equation for hot electron we
consider inertia term as zero, whereas in the
momentum equation of cold electrons we consider a
viscous term as its mobility is less and more
responsive to the viscous forces.

\begin{equation}\label{eq1}
\frac{\partial n_{h}}{\partial t}+\frac{\partial\left(n_{h} u_{h}\right)}{\partial x}=0
\end{equation}
\begin{equation}\label{eq2}
\frac{\partial n_{c}}{\partial t}+\frac{\partial\left(n_{c} u_{c}\right)}{\partial x}=0
\end{equation}
\begin{equation}\label{eq3}
0=\frac{e}{m_{e}} \frac{\partial \phi}{\partial x}-\frac{1}{m_en_{h}} \frac{\partial P_{h}}{\partial x}+\frac{\hbar^{2}}{2 m_{e}\gamma^2_h} \frac{\partial}{\partial x}\left[\frac{1}{\sqrt{n_{h}}} \frac{\partial^{2} \sqrt{n_{h}}}{\partial x^{2}}\right]
\end{equation}
This is momentum equation for hot electrons. The
inertia term (left hand side of the equation) is zero,
as hot electrons are very mobile so its inertia can be
assumed to be zero. Here, $\gamma_h= \left(1-\frac{u_h^2}{c^2} \right)^{-\frac{1}{2}}$
\begin{multline}\label{eq4}
\left(\frac{\partial}{\partial t}+u_{c} \frac{\partial}{\partial x}\right) (u_{c}\gamma_{c}) =\frac{e}{m_{e}} \frac{\partial \phi}{\partial x}\\
-\frac{1}{m_{e}n_{c}} \frac{\partial P_{c}}{\partial x} +\frac{\hbar^{2}}{2 m_{e}^{2}\gamma_{c}^{2}} \frac{\partial}{\partial x}\left[\frac{1}{\sqrt{n_{c}}} \frac{\partial^{2} \sqrt{n_{c}}}{\partial x^{2}}\right]\\
+\eta_c\frac{\delta^2u_c}{\delta x^2}
\end{multline}
This is momentum equation for cold electrons. The
inertia term is non zero as the mobility of cold
electrons is very less. For our two-electron plasma
system cold electron produces the restoring force for
electron-acoustic oscillations. Cold electrons also
got a viscous term. Here, $\gamma_c= \left(1-\frac{u_c^2}{c^2} \right)^{-\frac{1}{2}}$
\begin{equation}\label{eq5}
\frac{\partial^{2} \phi}{\partial x^{2}}=4 \pi e \left(n_{c}+n_{h}-Z_{i} n_{i}\right)
\end{equation}
The subscript $i$ in the Poisson’s equation is for
representing ions. The above-mentioned equations
are the governing equations for the plasma system,
where $n_h$ and $n_c$ are the density of hot and cold
electrons respectively. $\phi$ is the electrostatic
potential. $P_h$ is the pressure law for hot electrons. $\hbar$ is the plank constant divided by $2\pi$. $m_e$
is the mass
of electron and $e$ is the charge of electron and $\eta_c$
is the viscous constant for cold electrons.

We considered Fermi pressure as the main
pressure component for electrons. Fermi pressure is
\begin{equation}\label{eq6}
P_{j}=\frac{m_{j} V_{F j}^{2}}{3 n_{j 0}^{2}} n_{j}^{3} 
\end{equation}
where subscript $j=h$ is for hot electrons and $j=c$
for cold electrons. $n_{h0}$ is the initial hot electron density.
The parameters need to be normalized to get a good
control of the equations and normalized equations
are easy to work with. Normalization means we need
to define some suitable scaling associated with our
problem. With these scaling constants we can make
our parameters dimensionless.

For our problem, normalization has been done in
following ways
 $x \rightarrow x \omega_{j} / V_{F j}, t \rightarrow t \omega_{j}, \phi \rightarrow e \phi / 2 k_{B} T_{F j},  n_{h} \rightarrow n_{h} / n_{h 0}, n_{c} \rightarrow n_{c} / n_{c 0},, u_{h} \rightarrow u_{h} / V_{F h} , u_{c} \rightarrow u_{c} / V_{F c} $ where $\omega_{j}=\sqrt{4 \pi e n_{c 0} e^{2}/m_{e}}$, $V_{Fh}=\sqrt{\frac{2K_B T_{Fh}}{m_e}}$ and $H=\hbar \omega_{j} / 2 k_{B} T_{F j}$
 Changing the variables and parameters accordingly
we have Normalized Governing Equations.
\begin{equation}\label{eq7}
\frac{\partial n_{h}}{\partial t}+\frac{\partial\left(n_{h} u_{h}\right)}{\partial x}=0
\end{equation}
\begin{equation}\label{eq8}
\frac{\partial n_{c}}{\partial t}+\frac{\partial\left(n_{c} u_{c}\right)}{\partial x}=0
\end{equation}
\begin{equation}\label{eq9}
0=\frac{\partial \phi}{\partial x}-n_{h} \frac{\partial n_{h}}{\partial t}+\frac{H^{2}}{2\gamma_{h}^2} \frac{\partial}{\partial x}\left[\frac{1}{\sqrt{n_{h}}} \frac{\partial^{2} \sqrt{n_{h}}}{\partial x^{2}}\right]
\end{equation}

\begin{align}\label{eq10}
\left(\frac{\partial}{\partial t}+u_{c} \frac{\partial}{\partial x}\right) (u_{c}\gamma_c)&=\frac{\partial \phi}{\partial x}\\\nonumber&+
\frac{H^{2}}{2\gamma_{c}^2} \frac{\partial}{\partial x}\left[\frac{1}{\sqrt{n_{c}}} \frac{\partial^{2} \sqrt{n_{c}}}{\partial x^{2}}\right]\\ \nonumber&+\eta_c\frac{\delta^2u_c}{\delta x^2}
\end{align}

\begin{equation}\label{eq11}
\frac{\partial^{2}\phi}{\partial x^{2}}=n_{c}+\frac{n_{h}}{\delta}-\frac{\delta_{i}}{\delta} n_{i}
\end{equation}
where $\delta = \frac{n_{c0}}{n_{h0}}$, $\delta_1 = Z_i\frac{n_{i0}}{n_{h0}}$.
\section{Linear Dispersion Relation}\label{Linear}

\begin{figure}[ht]
{	\centering
	\includegraphics[width=3in,angle=0]{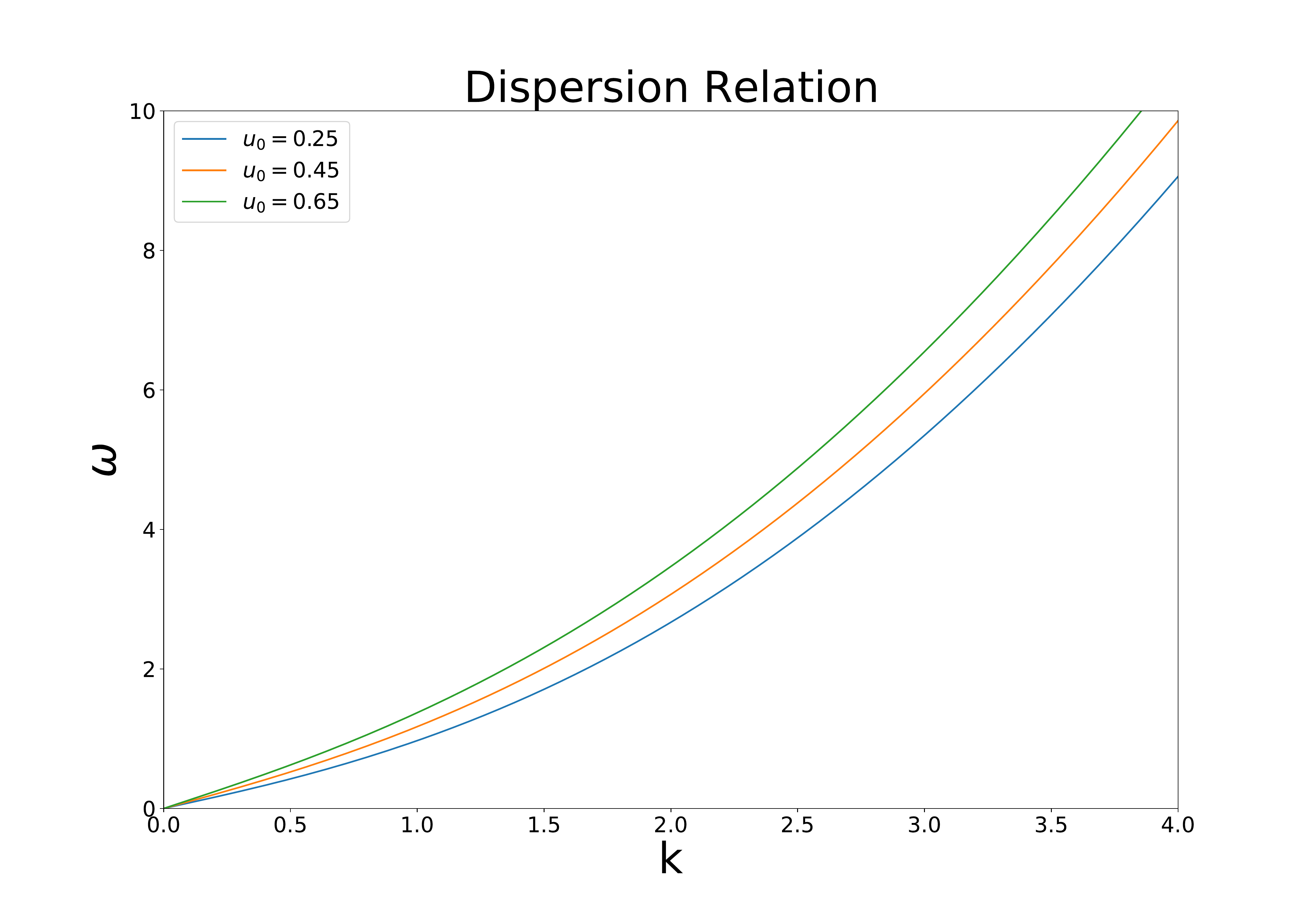}
	\caption{Dispersion relation plot for different $u_0$ keeping $\delta=0.3$ and $H=1$}
	\label{fig1}
}
\end{figure}

\begin{figure}[ht]
\centering
\includegraphics[width=3in,angle=0]{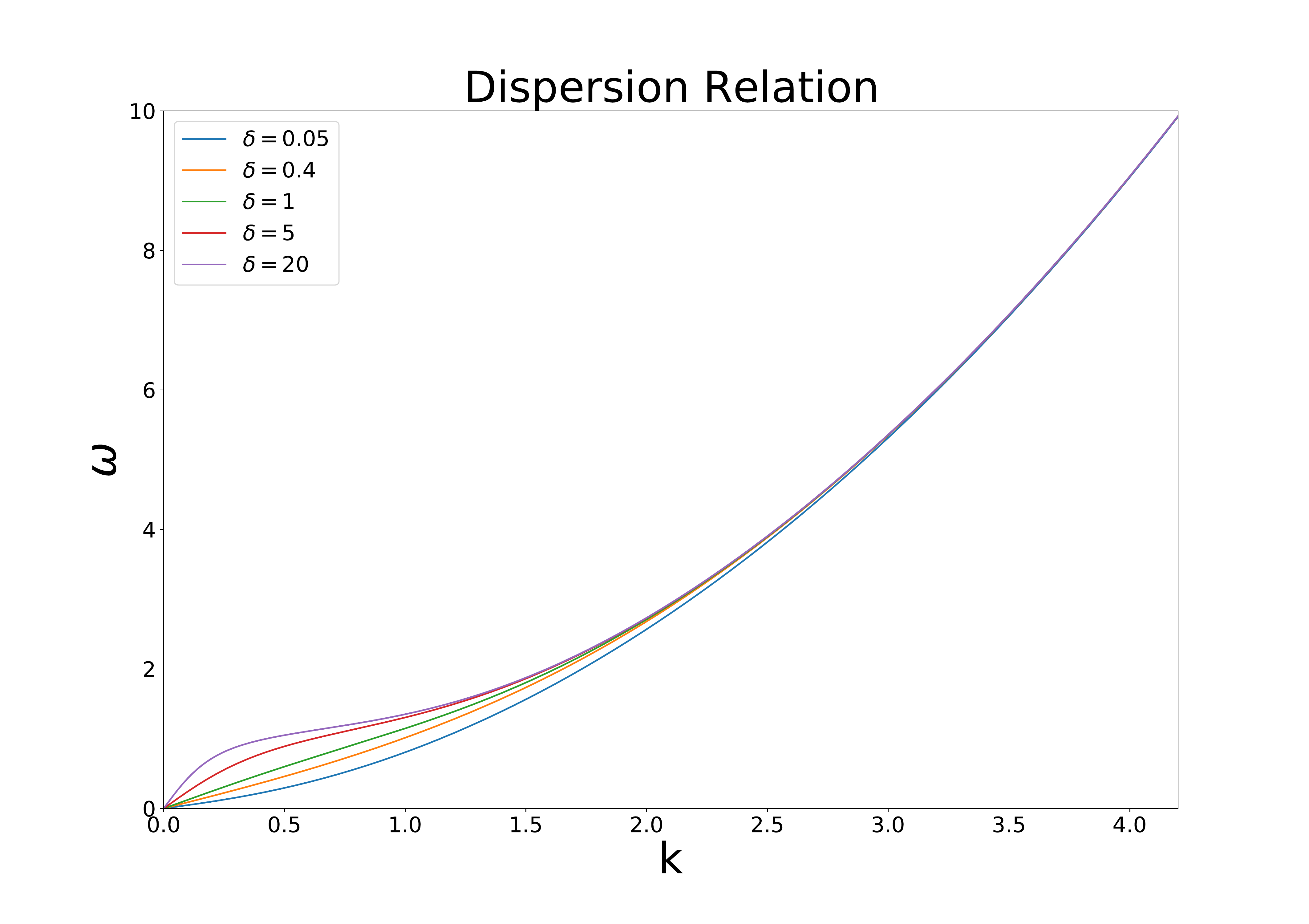}
\caption{Dispersion relation plot for different $\delta$ keeping $u_0=0.25$ and $H=1$}
	\label{fig2}
\end{figure}

\begin{figure}[ht]
{	\centering
	\includegraphics[width=3in,angle=0]{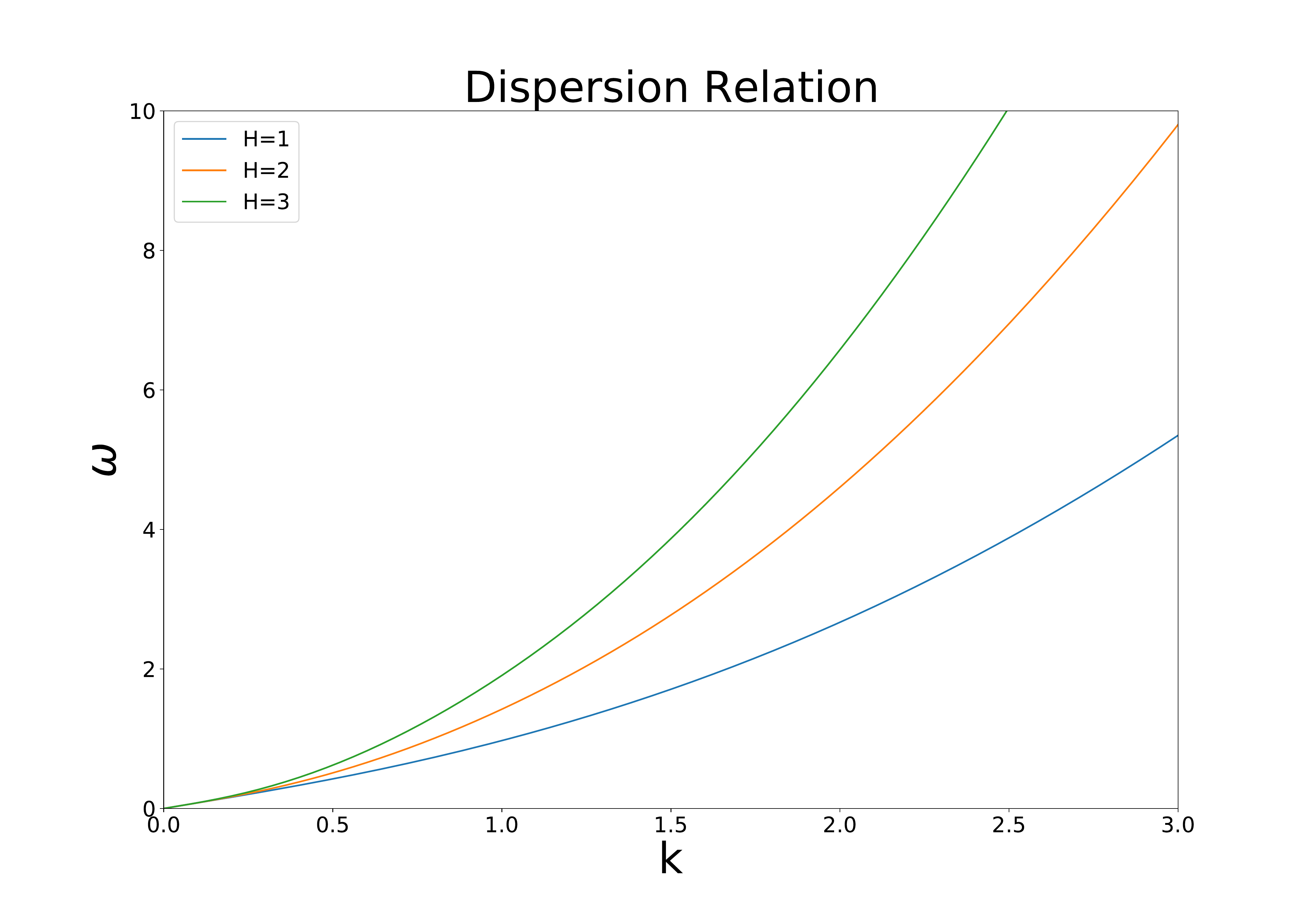}
	\caption{Dispersion relation plot for different $H$ keeping $\delta=0.3$ and $u_0=0.25$}
	\label{fig3}
}
\end{figure}

The system parameters can be expanded into
following perturbation expansion for $n_h$, $n_c$, $u_h$, $u_c$, and $\phi$
\begin{align}\label{eq12}
\left(\begin{array}{l}
n_{j} \\
u_{j} \\
\phi
\end{array}\right)=\left(\begin{array}{c}
1 \\
\pm u_{0 j} \\
\phi_{0}
\end{array}\right) &+\epsilon^{1}\left(\begin{array}{l}
n_{j}^{(1)} \\
u_{j}^{(1)} \\
\phi^{(1)}
\end{array}\right)+\epsilon^{2}\left(\begin{array}{l}
n_{j}^{(2)} \\
u_{j}^{(2)} \\
\phi^{(2)}
\end{array}\right) \\ \nonumber
&+\epsilon^{3}\left(\begin{array}{l}
n_{j}^{(3)} \\
u_{j}^{(3)} \\
\phi^{(3)}
\end{array}\right)+\cdots
\end{align}

Here we assume existence of streaming velocities for
hot and cold electrons (+ve for hot electron and -ve
for cold electrons) and equilibrium field in constant
field $\phi_0$. Substituting these expansions in the
governing relations and then taking only linear terms
(linearizing) with the assumption that all field
variable varies periodically as $e^{i(k x-\omega t)}$, we have following complex dispersion relation
\begin{multline}\label{eq13}
-k^{2}=\frac{1}{\left[\frac{H^{2} k^{2}}{4} \gamma_{c}^{2}-\left(\frac{\omega-u_{0} k}{k}\right)^{2}\right]+i \eta_{c}\left(\omega-u_{0} k\right)}\\
+\frac{\frac{1}{\delta}}{1+\frac{H^{2} k^{2}}{4} \gamma_{c}^{2}}
\end{multline}

Here we assume $u_{0 h}=u_{0 c}=u_{0} .$ Now we have to keep in mind that $k$ itself is a complex number, $k=$ $k_{1}+i k_{2}$. Putting this into the expression of Complex dispersion relation we have two equations, one for real part of the equation and other for imaginary part of the equation. The real dispersion equation is:
\begin{equation}\label{eq14}
-1=\frac{\left(\frac{1}{\delta}+k_{1}^{2}+\frac{H^{2} k_{1}^{4}}{4} \gamma_{c}^{2}\right) \cdot\left[\frac{H^{2} k_{1}^{4}}{4} \gamma_{c}^{2}-\omega^{2}-u_{0}^{2} k_{1}^{2}+2 \omega u_{0} k_{1}\right]}{\left[k_{1}^{2}+\frac{H^{2} k_{1}^{2}}{4} \gamma_{c}^{2}\right]}
\end{equation}

With $k_{2}=0,$ if we plot $\omega$ vs $k_{1}$ then the plot will give us dispersion plot. $H, u_{0}, \delta$ are the parameters of the equation.

Fig. \ref{fig1} shows dispersion plot for different values of equilibrium streaming velocity $u_{0} .$ Plot shows with increasing $u_{0}$ the slope of $\omega$ vs $k$ plot increases, i.e. an increase in EAW velocity. 

Fig. \ref{fig2}. shows dispersion plot for different values of
$\delta .$ Increase in $\delta$ shows decrease in slope (for small $\omega$ this behavior is clear). For $\delta=1$ i.e. the density of hot and cold electrons are same, we are getting a linear behaviour in small $k$ regime.

\begin{figure}
    \centering
    \includegraphics[width=3in,angle=0]{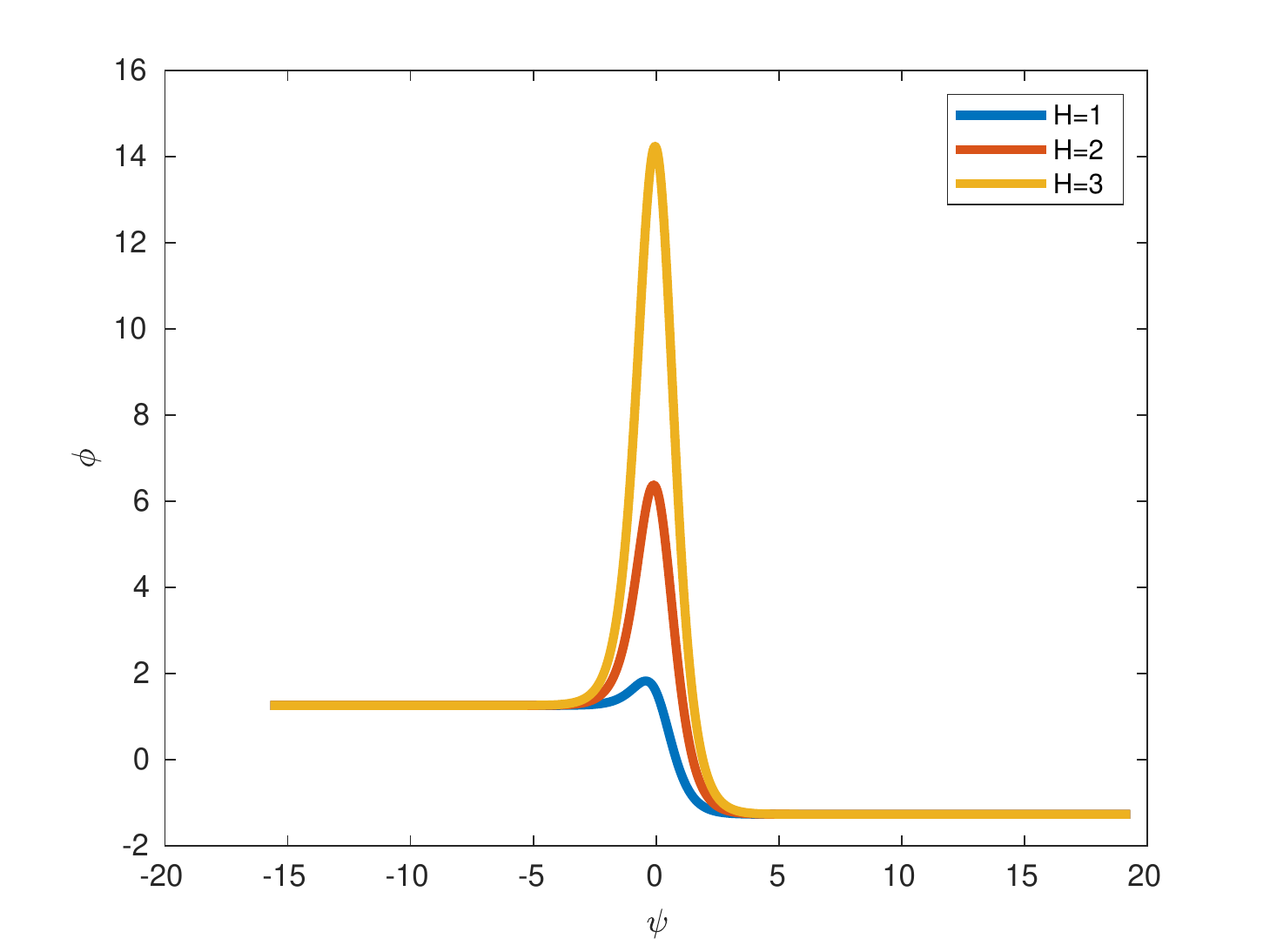}
    \caption{2D Solitary profile for different $H$ keeping $\delta$ and $u_0$ constant}
    \label{fig4}
\end{figure}
\begin{figure}
    \centering
    \includegraphics[width=3in,angle=0]{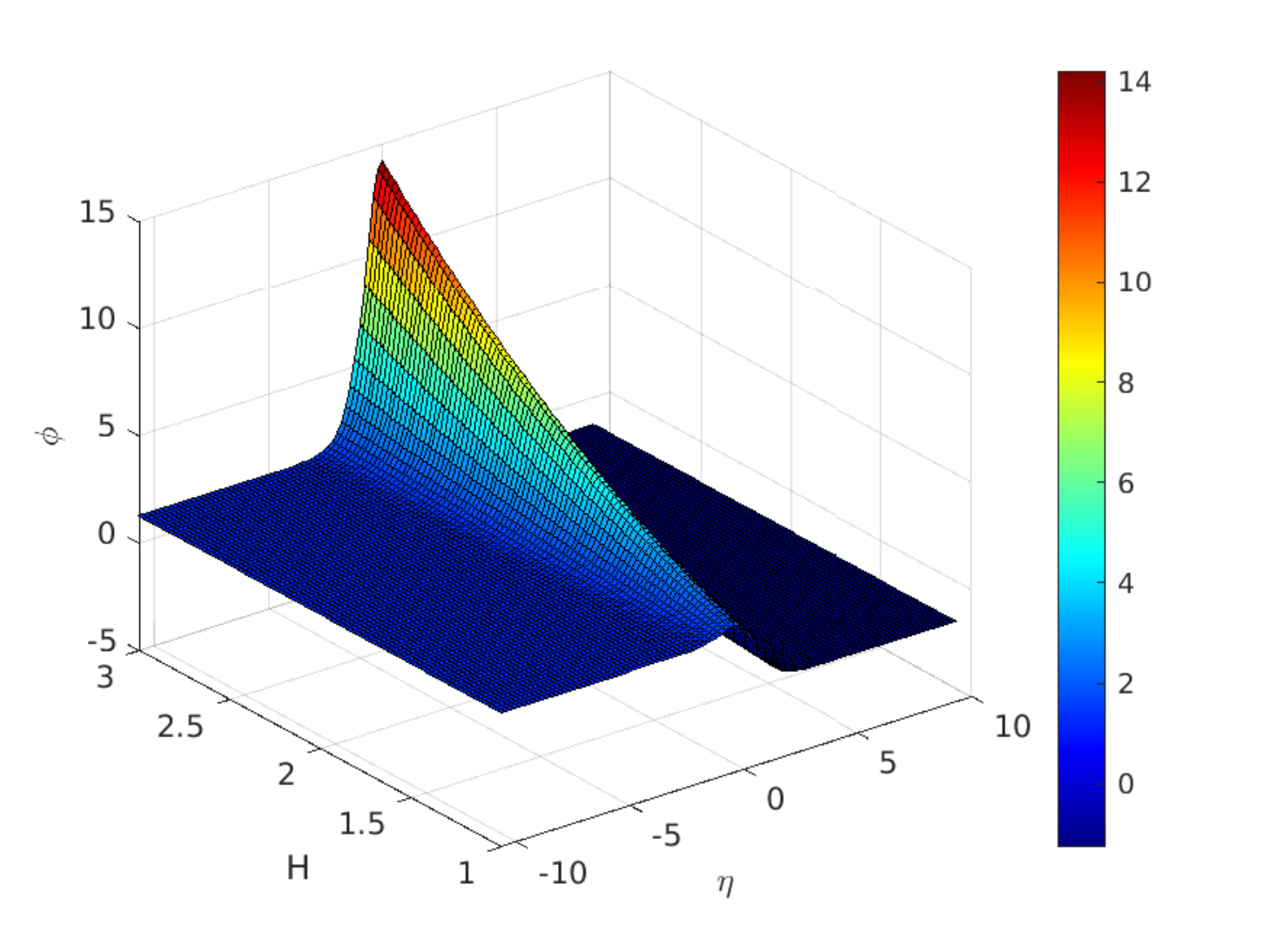}
    \caption{3D Solitary profile for different $H$ keeping $\delta$ and $u_0$ constant}
    \label{fig5}
\end{figure}
\begin{figure}
    \centering
    \includegraphics[width=3in,angle=0]{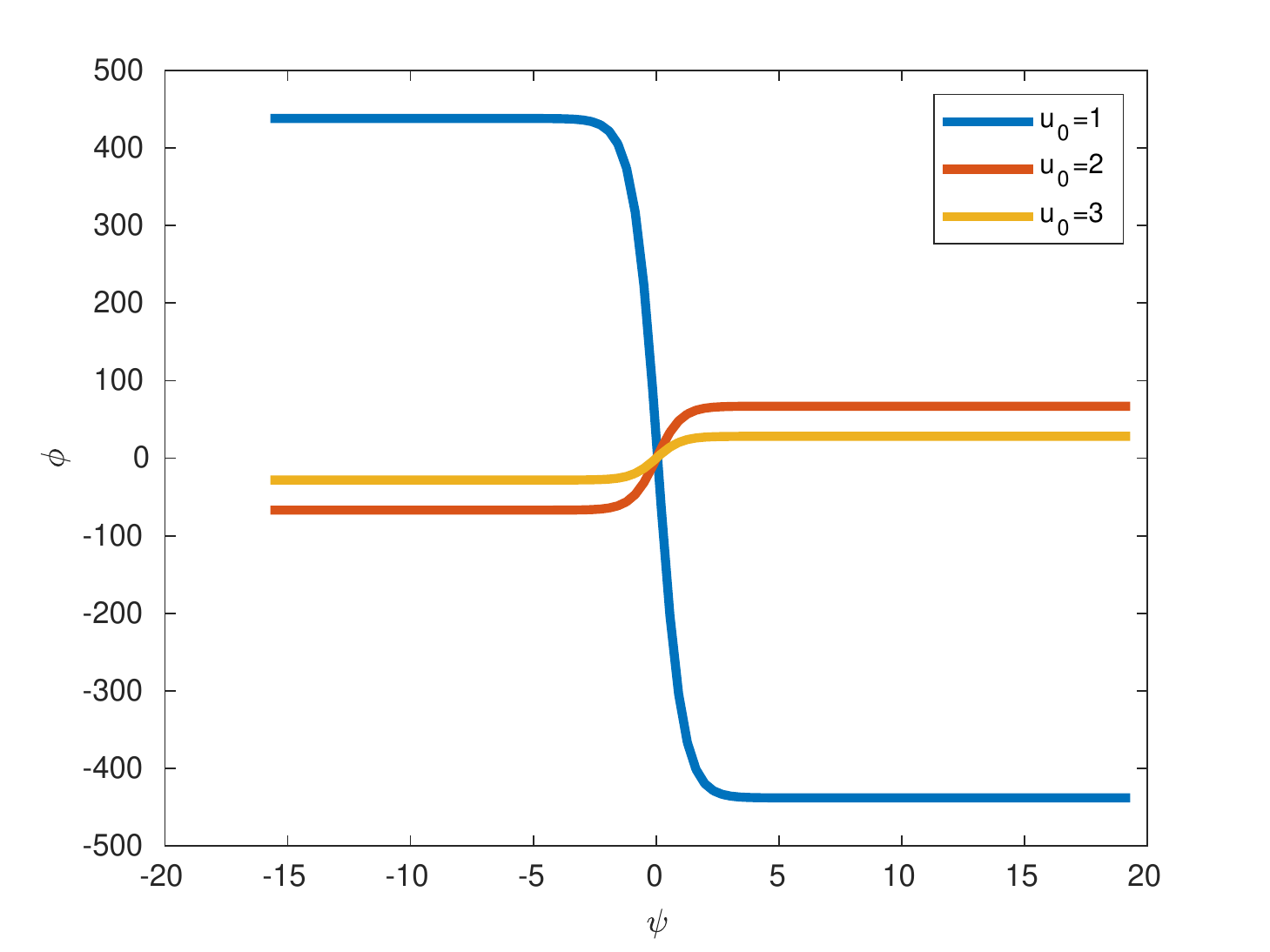}
    \caption{2D Solitary profile for different $u_0$ keeping $\delta$ and $H$ constant}
    \label{fig6}
\end{figure}
\begin{figure}
    \centering
    \includegraphics[width=3in,angle=0]{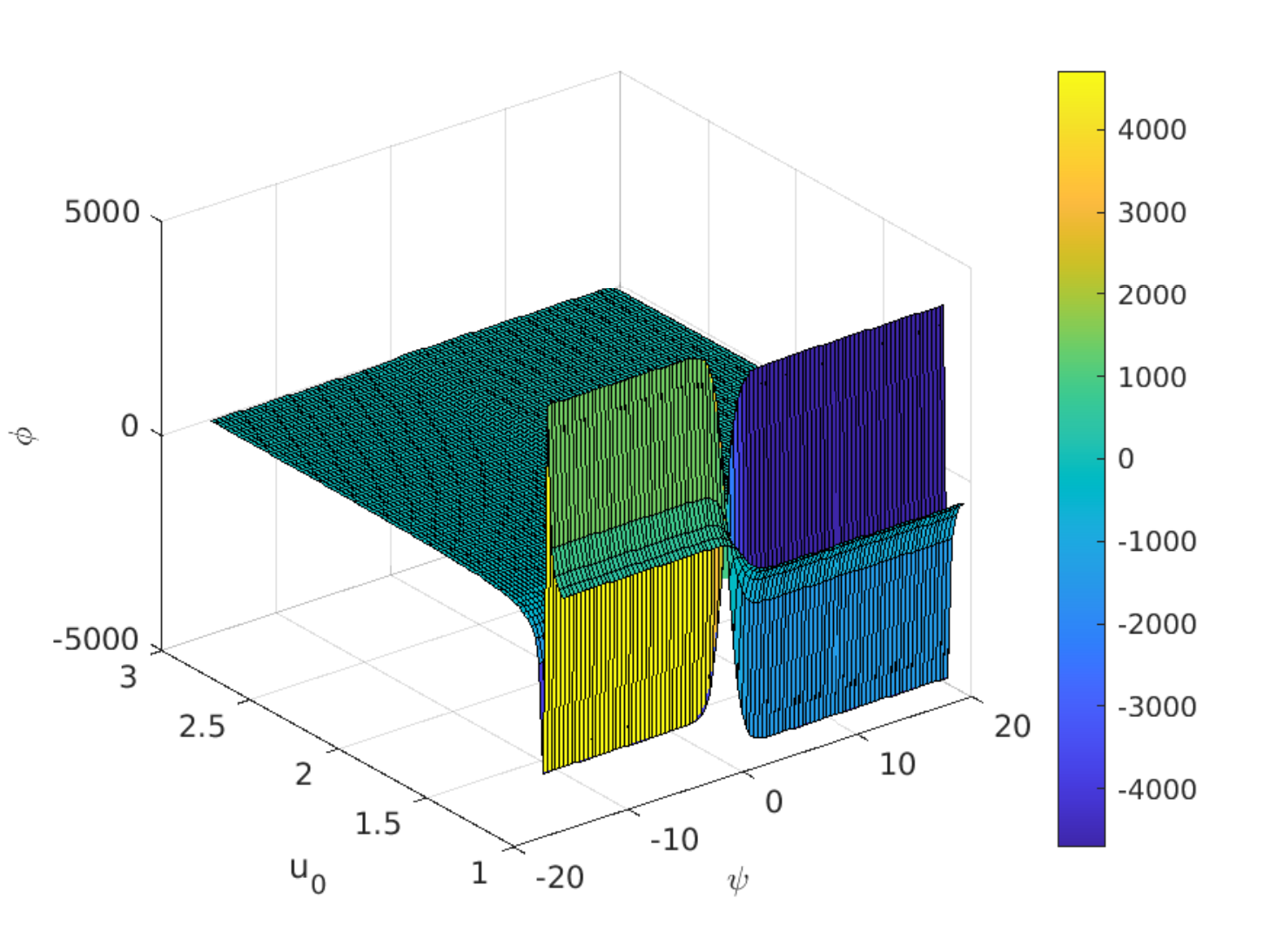}
    \caption{3D Solitary profile for different $u_0$ keeping $\delta$ and $H$ constant}
    \label{fig7}
\end{figure}
\begin{figure}
    \centering
    \includegraphics[width=3in,angle=0]{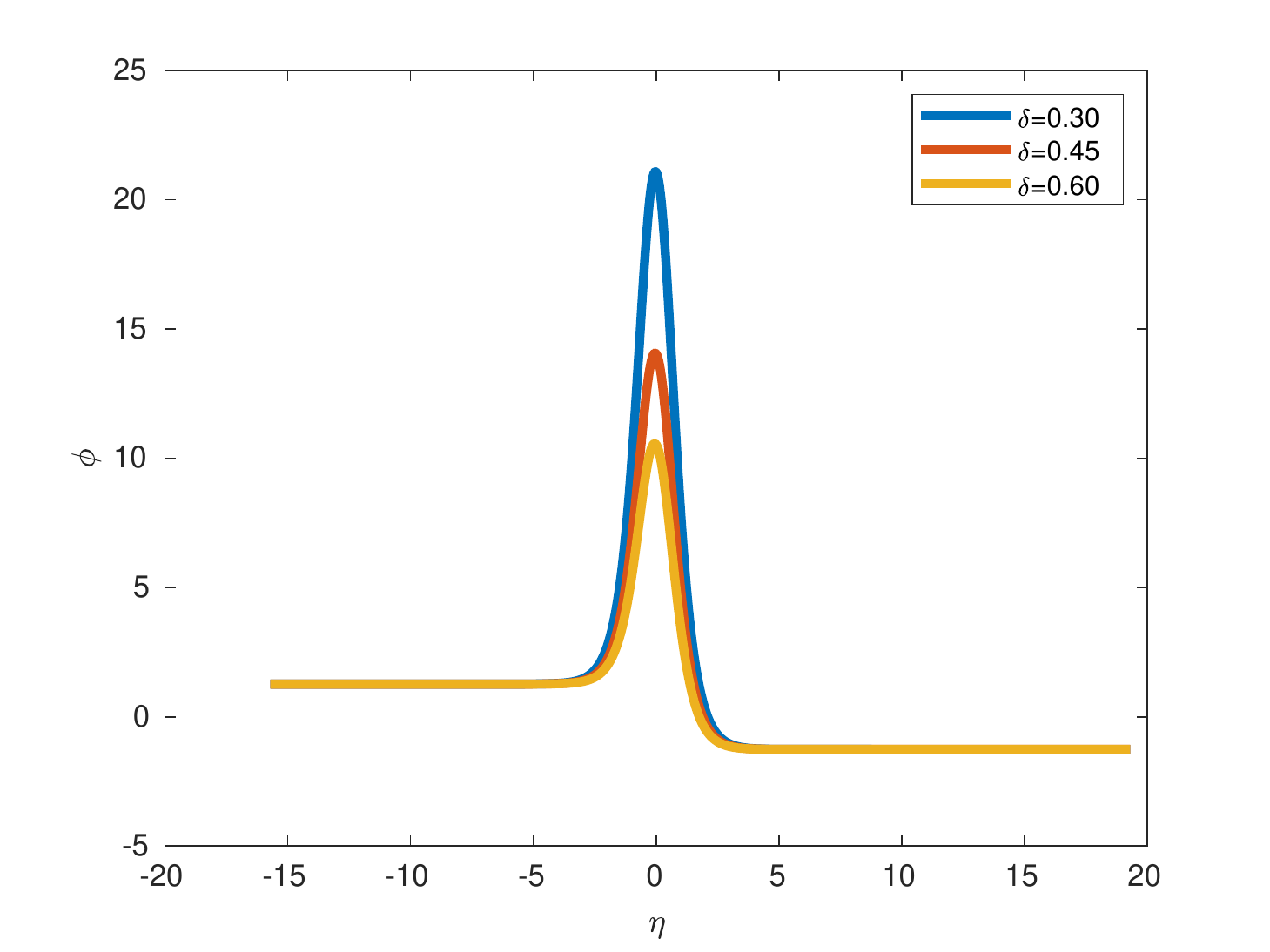}
    \caption{2D Solitary profile for different $\delta$ keeping $H$ and $u_0$ constant}
    \label{fig8}
\end{figure}
\begin{figure}
    \centering
    \includegraphics[width=3in,angle=0]{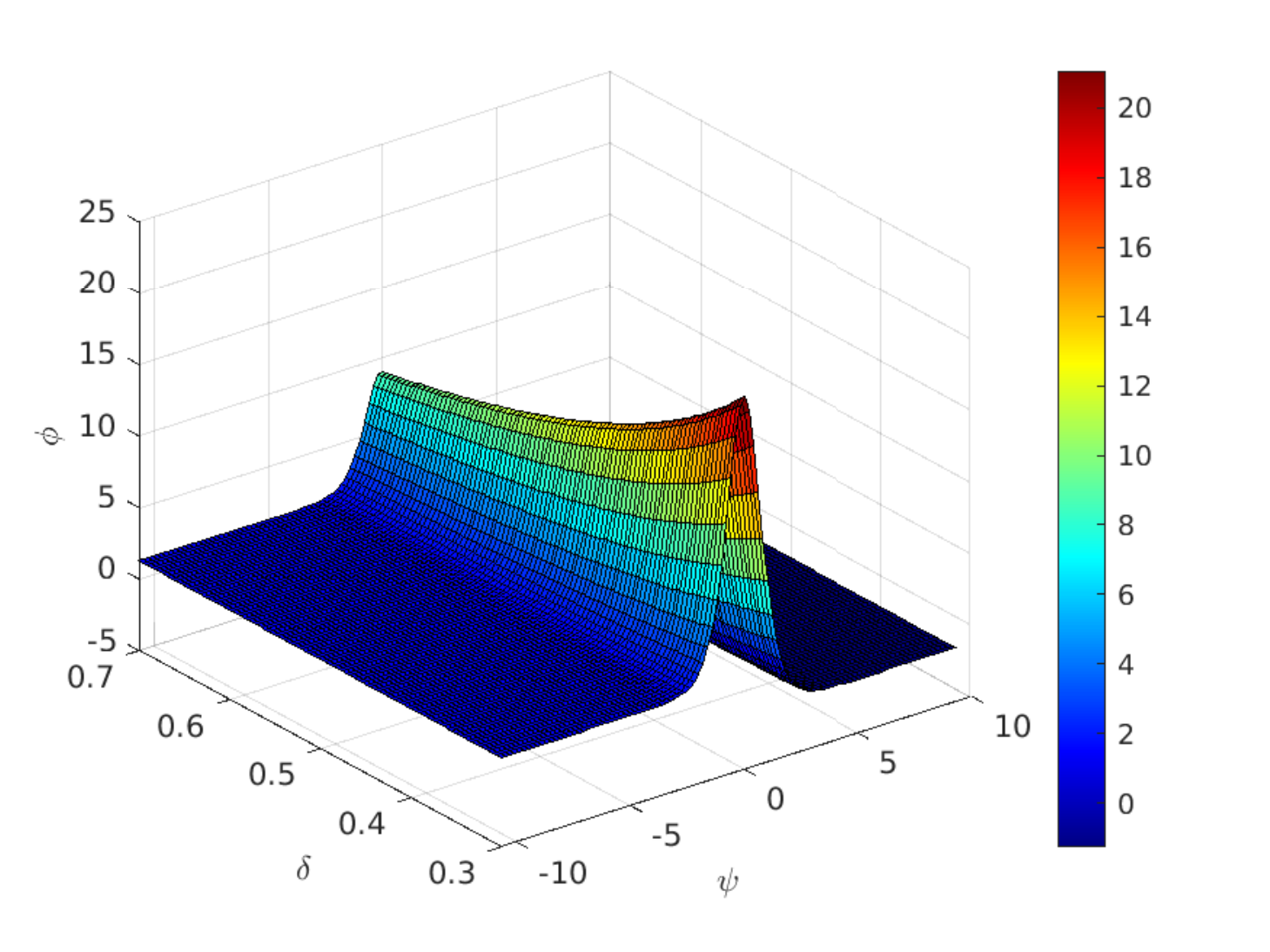}
    \caption{3D Solitary profile for different $\delta$ keeping $H$ and $u_0$ constant}
    \label{fig9}
\end{figure}

Increasing delta means more initial cold electrons compare to hot electrons. So, according to our plots a greater number of initial cold electrons implies a decrease in $k$ vs $\omega$ slop, that is increase in EAW velocity. Velocity of sound waves, rather acoustic waves increases with increasing the rigidity of the medium. Now a greater number of cold electrons means more viscous drag and eventually more rigid medium, which can be the cause for increase in EAW velocity.

Fig. \ref{fig3}. shows $\omega$ vs $k$ plot for different $H$ values. We change $H$ from 1 to 3 keeping $u_{0}, \delta$ and constant. As $H$ increases the slope of $\omega$ vs $k$ plot increases, that is also visible from the dispersion relation. However, $H$ depends on plasma frequency $\omega_{c}$ linearly. So, higher $H$ means higher plasma frequency. This leads
to a conclusion that for high plasma frequency the
EAW has less wavenumber with same frequency.

\section{Korteweg-de Vries Burgers Equation
for Nonlinear Study}\label{KDVB} We used standard reductive perturbation technique with usual stretching of space and time variables. We used two perturbation expansions to see the behavior of the solution of KdV-B equation. Using the
perturbation expansion eq. (\ref{eq12}) and, stretching of
variables
\begin{equation}\label{eq15}
\xi=\epsilon^{\frac{1}{2}}\left(x-v_{0} t\right) ; \quad \eta=\eta_{0} \epsilon^{\frac{1}{2}} ; \quad \text{and} \quad \tau=\epsilon^{\frac{3}{2}} t
\end{equation}
Putting them into governing equations and taking the lower order terms with power of epsilon we have the following KdV-Burger equation
\begin{equation}\label{eq16}
\frac{\partial \phi}{\partial \tau}+A \phi \frac{\partial \phi}{\partial \xi}+B \frac{\partial^{3} \phi}{\partial \xi^{3}}-C \frac{\partial^{2} \phi}{\partial \xi^{2}}=0
\end{equation}
The solution to (\ref{eq16}) is
\begin{equation}\label{eq17}
\phi=\frac{12 D}{N} \operatorname{sech}^{2} \xi-\frac{36 R}{15 N} \tanh \xi 
\end{equation}
with,
\begin{equation}\label{eq18}
A=\frac{2 D S_{1} R_{1}+F n_{0 c} R_{1}^{2}-D\left(M-u_{0 c}\right) \frac{Q_{1}^{2}}{n_{0 h} \delta}}{D S_{1}+n_{0 c} R_{1}\left(1+\frac{3 u_{0 c}^{2}}{2 c^{2}}\right)}
\end{equation}
\begin{equation}\label{eq19}
B=\frac{\left(D\left(M-u_{0 c}\right)\left(\frac{T_{h} Q_{1}}{n_{0} h^{\delta}}\right)-T_{c} S_{1} n_{0 c}-D\left(M-u_{0 c}\right)\right)}{D S_{1}+n_{0 c} R_{1}\left(1+\frac{3 u_{0 c}^{2}}{2 c^{2}}\right)}
\end{equation}
\begin{equation}\label{eq20}
C=\eta_{c} R_{1}
\end{equation}
With,
\begin{equation}\label{eq21}
D=M+\frac{3 M u_{0 c}^{2}}{2 c^{2}}-\frac{3 u_{0 c}^{3}}{2 c^{2}}-u_{0 c}
\end{equation}
\begin{equation}\label{eq22}
F=1+\frac{9 u_{0 c}^{2}}{2 c^{2}}-\frac{3 M u_{0 c}}{c^{2}}
\end{equation}
\begin{equation}\label{eq23}
T_{c}=\frac{H^{2}}{4 n_{o c}^{2}}\left(1-\frac{u_{0 c}^{2}}{c^{2}}\right)
\end{equation}
\begin{equation}\label{eq24}
T_{h}=\frac{H^{2}}{4 n_{0 h}^{2}}\left(1-\frac{u_{0 h}^{2}}{c^{2}}\right)
\end{equation}
\begin{equation}\label{eq25}
S_{1}=\frac{n_{0 c}}{\left.\left(M-u_{0 c}\right)\left(u_{0 c}+\frac{3 u_{0 c}^{3}}{2 c^{2}}\right)-M\left(1+\frac{3 u_{0 c}^{2}}{2 c^{2}}\right)\right]}
\end{equation}
\begin{equation}\label{eq26}
R_{1}=\frac{1}{\left(u_{0 c}+\frac{3 u_{0}^{3} c}{2 c^{2}}\right)-M\left(1+\frac{3 u_{0}^{2} c}{2 c^{2}}\right)}
\end{equation}
\begin{equation}\label{eq27}
Q_{1}=\frac{1}{n_{0 k}}
\end{equation}
$\text { For, } n_{0 j}=1 ; u_{0 j}=0 ; j=h c$

$S_{1}=-\frac{1}{M^{2}} ; R_{1}=-\frac{1}{M} ; Q_{1}=1 ; D=M ; F=1 ; T=\frac{H^2}{4}$

The the value of
\begin{equation}\label{eq28}
A=\left(\frac{M^{3}}{2 \delta}-\frac{3}{2 M}\right)
\end{equation}
\begin{equation}\label{eq29}
B=\left(\frac{H^{2}}{8 M}+\frac{M^{3}}{3}-\frac{M^{3} H^{2}}{4 \delta}\right)
\end{equation}
\begin{equation}\label{eq30}
C=-\frac{\eta_{0}}{M}
\end{equation}

FIG.\ref{fig4} and FIG.\ref{fig5} shows an increase in the peak of the solitary profile with increasing $H$ For this plot $u_{0}$=0.5 and $\delta$=0.45. FIG.\ref{fig7} and FIG.\ref{fig8} show decrease in the peak of the solitary profile with increasing $\delta$. For this plot $u_{0}$=0.5 and $H$=2

\section{Rogue wave}\label{Rogue}
KdV-B equation gives the solitary wave structures
which are created due to the balance between the
dispersive force and non-linear force in plasma. In
case the balance breaks the KdV-B starts to modify
itself to a sudden wave having amplitude almost
twice or greater than the amplitude of normal waves
at that time. Like the sudden high amplitude, the
suddenness of its timing makes it more dangerous.
The study of the Rogue wave is important as the
interference of the waves from different directions
and the non-linear property make it unpredictable.
We study the stability of Rogue wave and its
evolution from our KdV-B equation
\begin{equation}\label{31}
    F=\epsilon^{2} F_{0}+\sum_{s=1}^{\infty} \epsilon_{s}\left(F_{s} e^{i s \psi}+F_{s}^{*} e^{-i s \psi}\right)
\end{equation}
Where, $F$ is the field variable, is the smallness
parameter, and $\psi$ is the phase factor. Now, using this equation we can expand the field variable i.e. the potential in KdV-B equation
\begin{equation}\label{eq32}
    \phi=\epsilon^{2} \phi_{0}+\epsilon \phi_{1} e^{i \psi}+\epsilon \phi_{1}^{*} e^{-i \psi}+\epsilon^{2} \phi_{2} e^{2 i \psi}+\epsilon^{2} \phi_{2}^{*} e^{-2 i \psi}+\cdots
\end{equation}
The first harmonics ($\phi_1$) and second harmonics ($\phi_2$) can be further expanded respectively

\begin{figure}
    \centering
    \includegraphics[width=3in,angle=0]{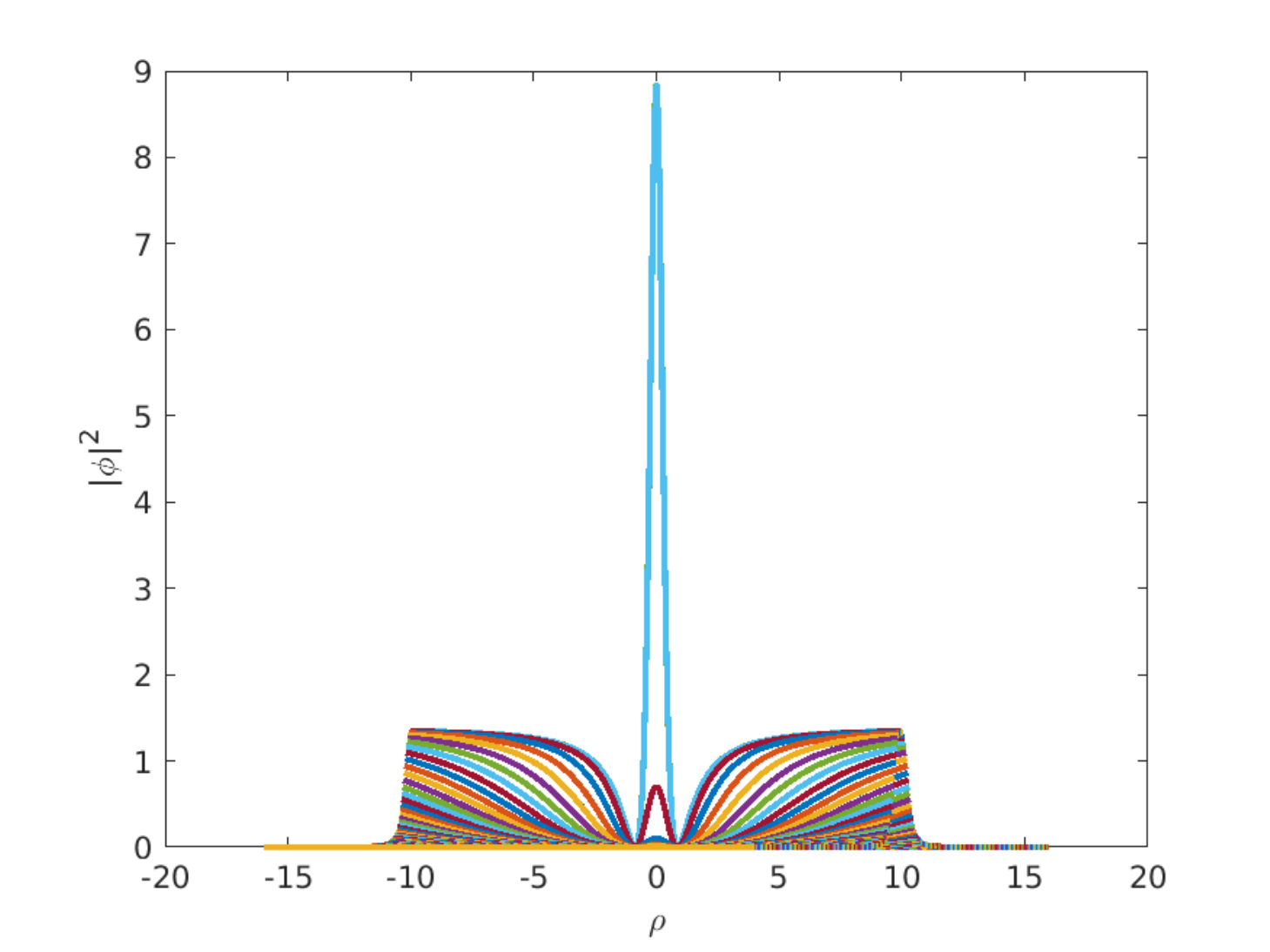}
    \caption{Rogue wave structure}
    \label{fig10}
\end{figure}

\begin{equation}
\phi_{1}=\phi_{1}^{(1)}+\epsilon \phi_{1}^{(2)}+\epsilon^{2} \phi_{1}^{(3)}+\ldots
\end{equation}
\begin{equation}
\phi_{2}=\phi_{2}^{(1)}+\epsilon \phi_{2}^{(2)}+\epsilon^{2} \phi_{2}^{(3)}+\ldots
\end{equation}

$F_0$ and $F_s$ are assumed to vary very slowly with space and time.
Now, we use the change in variables $\frac{\partial}{\partial \tau}=-i s \omega-\epsilon c \frac{\partial}{\partial \rho}+\epsilon^{2} \frac{\partial}{\partial \theta} \text { and } \frac{\partial}{\partial \xi}=i s k+\epsilon \frac{\partial}{\partial \rho}$ and put it in the Eq. (\ref{eq16}) and equate the coefficients of the first order of $\epsilon$, we get

\begin{equation}
\omega=-B k^{3}
\end{equation}
\begin{equation}
\frac{d \omega}{d k}=-3 B k^{2}
\end{equation}
Equating the coefficients of $e^{2i\psi}$ with $\epsilon^2$
\begin{equation}
\varphi_{2}^{(1)}=\frac{A}{6 B k^{2}} \varphi_{1}^{(1)^{2}}
\end{equation}
Equating the terms independent of $\psi$ with $\epsilon^3$
\begin{equation}
    \varphi_{0}^{(1)}=\frac{A}{C} \varphi_{1}^{(1)} \varphi_{1}^{(1)^{*}}
\end{equation}
By equating the other higher terms we get
\begin{equation}
\frac{\partial \varphi_{1}^{(1)}}{\partial \tau}+3 i k B \frac{\partial^{2} \varphi_{1}^{(1)}}{\partial \rho^{2}}=-i A k\left(\varphi_{0}^{(1)} \varphi_{1}^{(1)}+\varphi_{2}^{(1)} \varphi_{1}^{(1)^{*}}\right)
\end{equation}
\begin{equation}\label{eq39}
i \frac{\partial \varphi_{1}^{(1)}}{\partial \tau}-3 k B \frac{\partial^{2} \varphi_{1}^{(1)}}{\partial \rho^{2}}+\frac{A^{2}}{6 B k}\left(\varphi_{1}^{(1)^{2}} \varphi_{1}^{(1)^{*}}\right)=0
\end{equation}
Let's assume $P=i c-3 B k$ and
$Q=A^{2} k\left[\frac{1}{6 B k^{2}+4 i k c}-\frac{1}{3 B k^{2}}\right] .$ The term $\mathrm{PQ}$ is very
important in defining the existence of a rogue wave. In the region $P Q<0$ the rogue wave is stable and for $P Q>0$ the rogue wave doesn't exist. The solution of Eq. (\ref{eq39}) we get as
\begin{equation}
    \phi(\rho, \theta)=\sqrt{\frac{2 P}{Q}}\left[\frac{4(1+4 i P Q)}{1+16 P^{2} Q^{2}+4 \rho^{2}}-1\right] \exp (2 i P \theta)
\end{equation}
Here we plot the potential amplitude with the changed variable $\rho$ for different values of $\theta$. And we take almost 500 values of $\theta$ and as one can see that for only one value the potential suddenly gets abnormally high with respect to the other. All of the solutions we get from the same NLSE, where only one wave behaves dangerously. It perfectly depicts the feature of a Rogue Wave, the amplitude is way higher than that of solitary profiles in Fig. (\ref{fig4}) to Fig. (\ref{fig9}).

Here we take the viscosity coefficient $\eta_c=5$ and
observed that the amplitude of the rogue wave in the
plot is lower than that of the rogue wave with no
viscous term.

\section{Simulation of KdV-B and Rouge wave
evolution}\label{Simulation}
The static plots of two simulations of the solitary profiles and rogue wave profile of the relativistic plasma are shown below.

\begin{figure}[h]
    \centering
    \includegraphics[width=3in,angle=0]{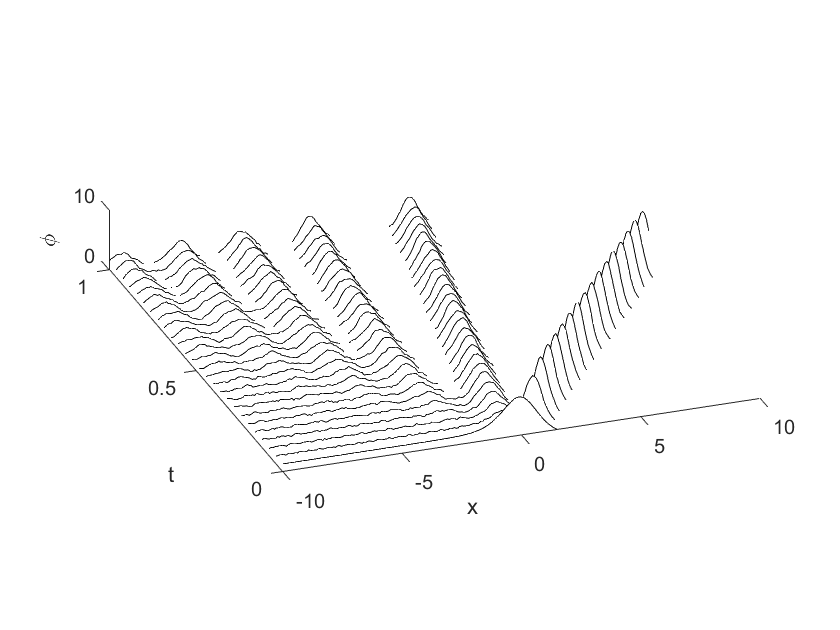}
    \caption{Simulation result of evolution of solitary profile}
    \label{fig11}
\end{figure}
\begin{figure}[h]
    \centering
    \includegraphics[width=3in,angle=0]{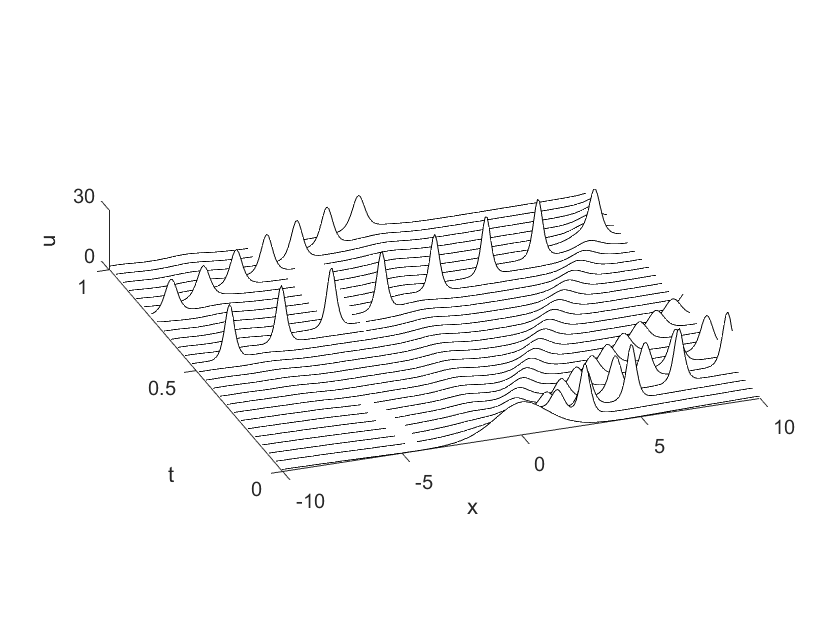}
    \caption{Simulation result of evolution of Rogue wave profiles}
    \label{fig12}
\end{figure}

A movie of KdV-B evolution can be found in this
link: \url{https://doi.org/10.5281/zenodo.4366748} and a
movie of Rouge wave evolution can be found in this
link: \url{https://doi.org/10.5281/zenodo.4384499} .

In both FIG.\ref{fig11} and FIG.\ref{fig12}, using the solutions of KdV-B and
rouge wave, we simulate the KDV solitary wave and
rogue wave respectively. Here, $‘x’$ represents the space coordinate and $‘t’$ represents the time
coordinate, and $‘u’$ is the amplitude. In both cases,
we see the changes in the waves in the interval
between $t = 0s$ and $t = 1s$. As you can see from the
simulations that rogue waves have much higher
amplitudes than those of normal solitary waves and
also less frequent than normal waves.

\section{Dynamical Study}\label{Dynamic}
To get the phase diagram of EAWs, we study a 
travelling wave solution followed by the
KdV-B equation. To do that we use the transformation $\eta=$ $\xi-M \tau$ with the proper boundary conditions at $\eta \rightarrow$ $\pm \infty, \psi, \delta \psi / \delta \eta$ and $\left(\delta^{\wedge} 2 \psi\right) /\left(\delta \eta^{\wedge} 2\right) \rightarrow 0$ where
$M$ is the velocity of the wave. Applying a transformation like \cite{ref56} we can write the Non-linear Schrodinger equation as

\begin{figure}[h]
    \centering
    \includegraphics[width=3in,angle=0]{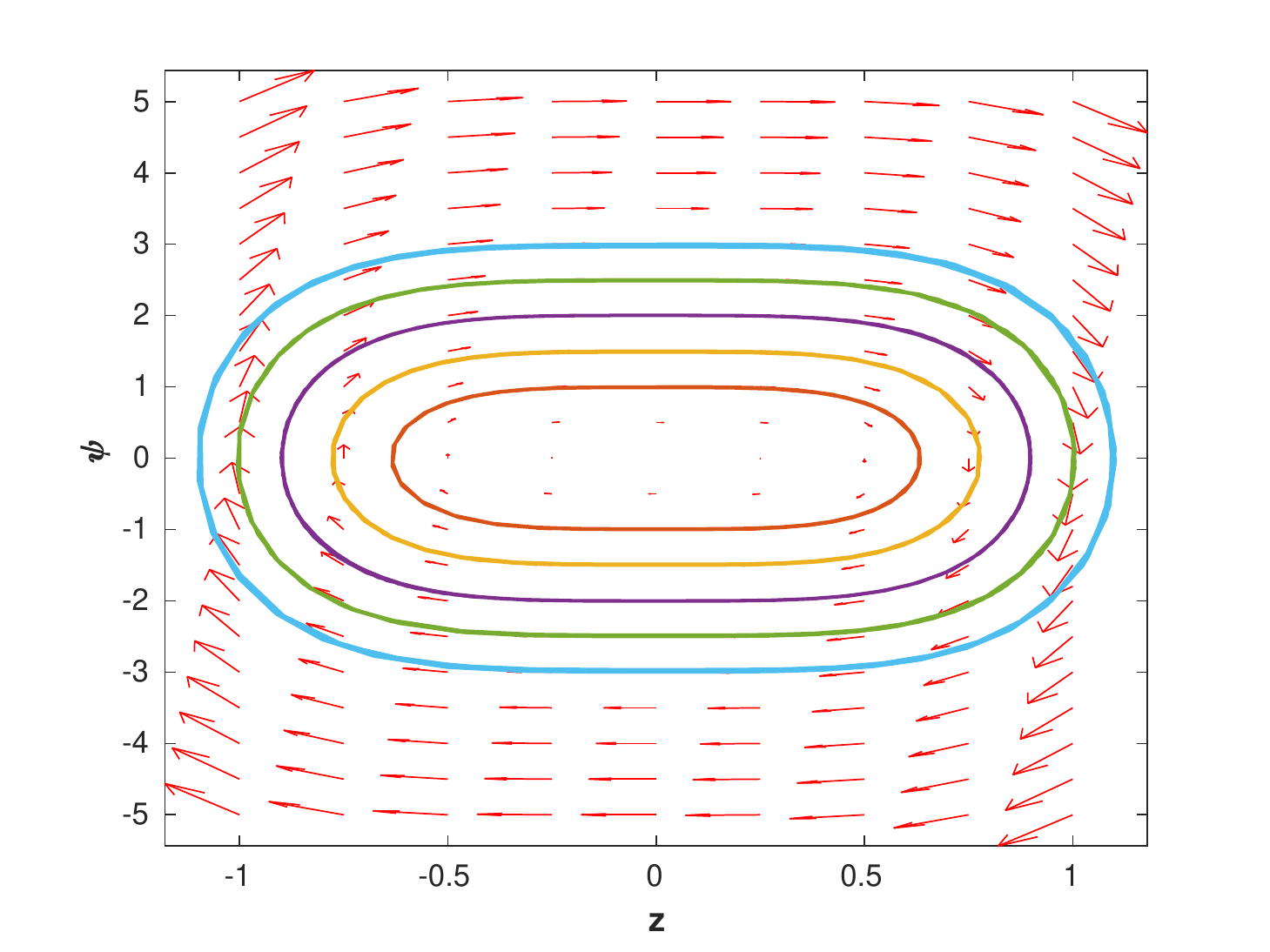}
    \caption{$\psi$ vs $z$ plot (Phase plot) for dynamical behaviour.}
    \label{fig13}
\end{figure}

\begin{equation}
\frac{d^{2} \psi}{d \eta^{2}}=\left(\beta^{2}-\frac{1}{M_{1} l^{2}} \beta V\right) \psi-\frac{M_{2}}{M_{1} l^{2}}
\end{equation}
Eq. (41) can be written in the following dynamical
form
\begin{equation}
\frac{d \psi}{d \eta}=z
\end{equation}
\begin{equation}
\frac{d z}{d \eta}=M_{1} \psi-M_{2} \psi^{3}
\end{equation}

\begin{figure}[ht]
    \centering
    \includegraphics[width=3in,angle=0]{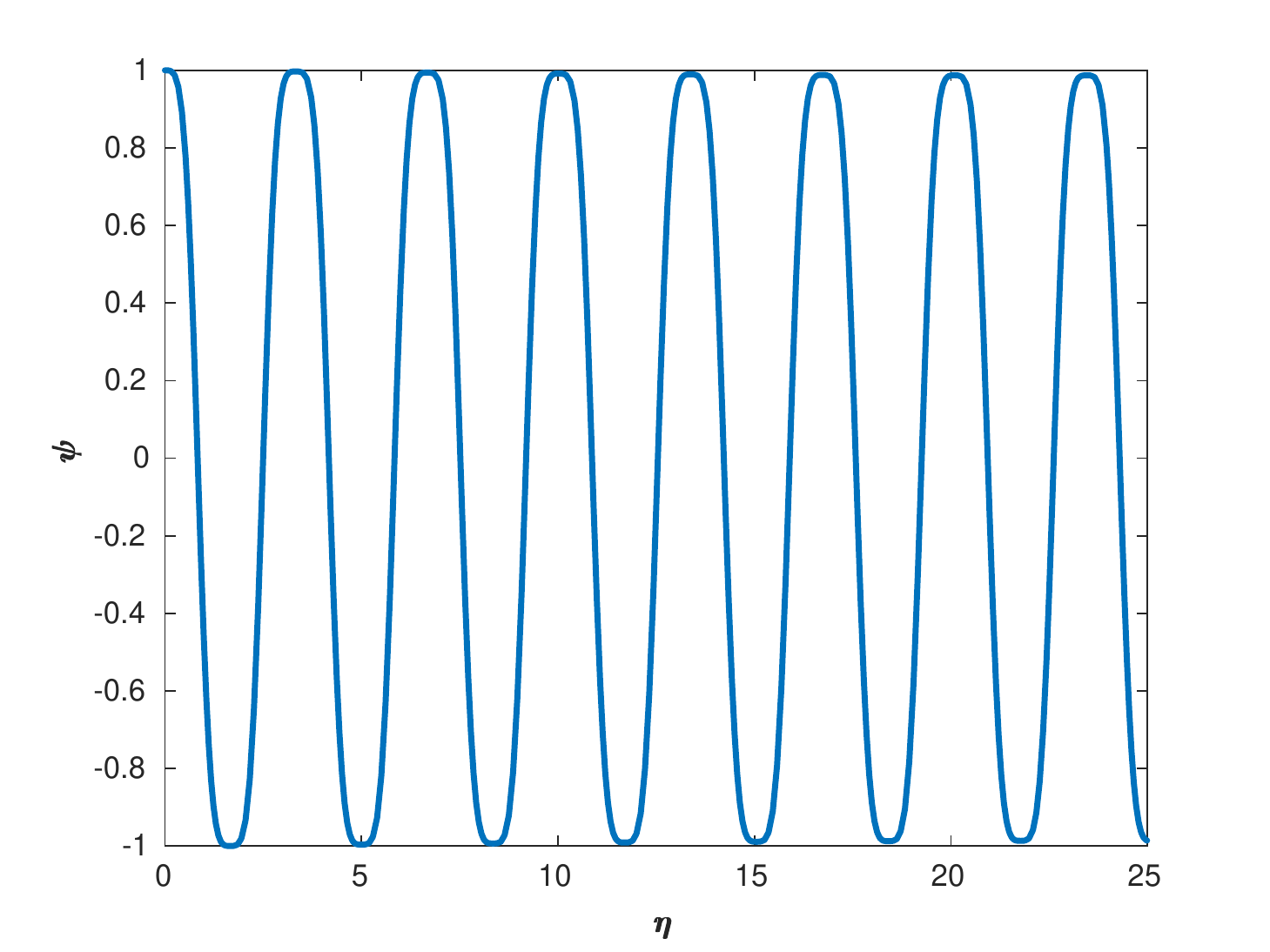}
    \caption{$\psi$ vs $\eta$ plot for dynamical behaviour.}
    \label{fig14}
\end{figure}

The phase plot gives the dynamical nature of the system (Fig.\ref{fig13}) The progressive nature of EAW can be seen from Fig.\ref{fig14}. Plot shows a stable solution. Different coloured plots in phase diagram correspond to different energy solutions. Green one is the highest energy and blue one is the lowest.


\section{summary}\label{summary}
Both the linear and nonlinear properties of electron-acoustic wave (EAW) have been investigated in three component Fermi plasma consisting of two of Fermi pressure for hot electrons. The stable models for EAWs are shown. The dependence of
electron-acoustic wave velocity on different plasma parameters is shown and explained. The slope of the dispersion relation plots depends on the velocity of
EAW inversely. The dependence of this EAW velocity and EAW frequency on different plasma parameters has been shown. Electron-acoustic wave velocity increases with increasing $H$, $u_0$ and $\delta$ and decreases with increasing viscous coefficient $\eta_c$ The formation of rouge wave in the system has been studied. The evolution of solitary profile and rouge wave has been studied. Dynamical properties of the system has also been studied and found a stable dynamical structure.

\begin{acknowledgments}
SP would like to acknowledge the support of institute fellowship by IIT Kharagpur and the scholarship provided by DST-INSPIRE. SC would like to thank the Institute of Natural Sciences and Applied Technology, the Physics departments of Jadavpur University and Government General Degree College at Kushmandi for providing facilities to carry this work.
\end{acknowledgments}


\bibliography{suman}

\end{document}